\ifdefined\WORDCOUNT
\documentclass[superscriptaddress,twocolumn,preprintnumbers,amsmath,amssymb,prl,nofootinbib,groupedaddress]{revtex4-1}
\else
\documentclass[superscriptaddress,twocolumn,showpacs,preprintnumbers,amsmath,amssymb,prl,groupedaddress]{revtex4-1}
\fi
\usepackage[table]{xcolor}
\usepackage{slashed}
\usepackage{cancel}

\usepackage{multirow}
\usepackage[hidelinks]{hyperref}

\newcolumntype{s}{>{\columncolor[HTML]{AAACED}} p{3cm}}

\usepackage{amssymb}
\usepackage{epsfig,amsmath,graphics}
\usepackage{epstopdf}
\usepackage{graphicx}
\usepackage{color}
\usepackage{gensymb}

\usepackage[utf8x]{inputenc}
\newcommand{\ket}[1]{\ensuremath{\left|#1\right>}}
\newcommand{\bra}[1]{\ensuremath{\left<#1\right|}}
\newcommand{\braket}[1]{\ensuremath{\left<#1\right>}}

\usepackage{natbib}
\usepackage{graphicx}

\begin{document}

\title{Proposal for matter-wave interferometry with a rare-earth-doped microparticle}

\author{Chris Overstreet}
\email{c.overstreet@jhu.edu}
\affiliation{Department of Physics and Astronomy, The Johns Hopkins University, Baltimore, Maryland 21218}

\begin{abstract} 
Matter-wave interferometers are sensitive probes of low energy physics and have been used for precise tests of gravity and quantum mechanics.  These applications would benefit from interfering particles of higher mass, but observing the interference of a large particle is challenging due to the need to control its initial state and to avoid decoherence.  
Here we propose to demonstrate matter-wave interference of a microparticle with an embedded rare-earth ion.  Optical transitions of the rare-earth ion will impart momentum to the microparticle's center of mass.  The use of a time-symmetric interferometer geometry and a rare-earth ion state with angular momentum $J = 1/2$ will eliminate sensitivity to the initial conditions of the microparticle.  We show that the rates of all relevant external decoherence mechanisms either decrease or remain constant as the microparticle mass is increased, allowing decoherence to be avoided by using sufficiently large particles.  An apparatus at moderate vacuum levels will support microparticle interferometry with up to $10^3$ photons per beam splitter and millisecond coherence time.  This demonstration will establish microparticle interferometry as a new platform for quantum sensing, improving searches for minimal modifications of quantum mechanics by up to three orders of magnitude in the near term and laying the foundation for future gravitational tests.

\end{abstract}
\maketitle

\section{I.  Introduction}  
\label{sec:Introduction}

Matter-wave interferometers \cite{Hogan2009, Overstreet2020a}, which delocalize and interfere the center-of-mass position of a massive particle, are the basis of many low-energy precision measurements and tests of fundamental physics \cite{Morel2020, Parker2018, Asenbaum2020a}.  In particular, acceleration-sensitive interferometers have been used to test the equivalence principle of general relativity at the $10^{-12}$ level \cite{Asenbaum2020a}, to observe gravitational phenomena in quantum systems \cite{Overstreet2022, Asenbaum2017}, and to constrain minimal modifications of quantum mechanics \cite{nimmrichter2013macroscopicity, schrinski2020quantum, zhu2026high}.  They have also been proposed for dark matter searches \cite{Abe2021, Zhou2024}, for gravitational wave detection \cite{Abe2021, Graham2013}, and to search for gravitationally-mediated entanglement generation \cite{Bose2017,Marletto2017}.  

All of these applications would benefit from the ability to interfere particles of much higher mass than has previously been demonstrated.  In Mach-Zehnder accelerometers, for example, the phase sensitivity to acceleration is given by $\phi/g = k T^2$, where $\phi$ is the interferometer phase, $g$ is the relative acceleration between the interfering particles and the beam splitter, $\hbar k$ is the momentum imparted by the beam splitter, and $T$ is the interferometer time.  The wave packet separation between the two interferometer trajectories is given by $\Delta x = \hbar k T/m$, where $m$ is the mass of the interfering particle.  Since both trajectories must remain within the apparatus, the length scale of the apparatus restricts $\Delta x$ and thus, for a given $T$, imposes a constraint on $k$ and $\phi/g$ \cite{footnote1}.  
This constraint can be loosened by increasing $m$. 
Likewise, for experiments aiming to detect gravitationally mediated entanglement generation to succeed, the mass of the interfering particles must be $\sim 10^{12}$ amu \cite{Bose2017}, which is a factor of $10^7$ higher than the current state of the art \cite{pedalino2026probing}.  There is thus great interest in developing a system that enables the quantum interference of large particles, with many recent proposals along these lines \cite{delic2020cooling, delord2020spin, scala2013matter, kaltenbaek2023research, bose2025}. 

Three key criteria must be satisfied for the matter-wave interference of a large particle to be observed.  First, the system must implement a beam splitter that can delocalize and interfere the center-of-mass position of the large particle.  Second, the measured quantity must be insensitive to the initial conditions of the large particle, including its center-of-mass position and momentum, its orientation, its rotation rate, and its internal state.  This criterion removes the necessity of controlling these degrees of freedom of the large particle, which is difficult in practice.  Third, the system must avoid sources of decoherence, including background gas scattering, blackbody radiation scattering, and any decoherence mechanisms associated with the beam splitter.  To our knowledge, no previous proposal to observe the interference of a  microparticle meets all these criteria.  

In this work, we propose an experiment to demonstrate matter-wave interferometry with a microparticle containing a rare-earth ion.  Optical transitions between $4f$ states of the rare-earth ion will coherently transfer momentum to the microparticle's center of mass.  A three-pulse sequence will be used to implement a Mach-Zehnder interferometer, and the interferometer phase will be measured by detecting the quantum state of the rare-earth ion after the interferometer.  This approach combines a highly coherent beam splitter, insensitivity to the initial state of the microparticle, and the ability to mitigate all known decoherence mechanisms.    

The remainder of this paper is organized as follows:  Section~\hyperref[sec:Concept]{II} describes the experimental concept and discusses how the system satisfies the three criteria outlined above; Section~\hyperref[sec:Implementation]{III} describes an implementation using a samarium ion embedded in a SrSe microparticle; Section~\hyperref[sec:Decoherence]{IV} contains calculations of decoherence rates, and Section~\hyperref[sec:Discussion]{V} discusses prospects to use microparticle interferometry for tests of fundamental physics.  

\section{II.  Experimental concept}
\label{sec:Concept}

The idea of the experiment is to implement a Mach-Zehnder matter-wave interferometer by driving a sequence of optical Raman transitions in a rare-earth ion embedded in a microparticle.  If the Rabi frequency of the Raman transitions is below the frequency of the lowest phonon mode in the microparticle, the rare-earth ion is in the tight-binding (Lamb-Dicke) regime, and the momentum from each optical interaction is transferred to the microparticle center of mass.  The rare earth ion remains in a level with total angular momentum $J = 1/2$ throughout the interferometer, and its angular momentum projection quantum number is used for state readout once the interferometer is complete.  The concept is illustrated in Fig. 1.  We note that optical \cite{bottger2009effects} and electron spin \cite{siyushev2014coherent} coherence times exceeding $1$~ms have been observed in systems of rare-earth ions in solid-state materials, indicating that quantum control of rare-earth ions can be maintained at this time scale in the presence of internal decoherence mechanisms. 

\begin{figure}[h!]
    \centering
    \includegraphics[width=0.475\textwidth]{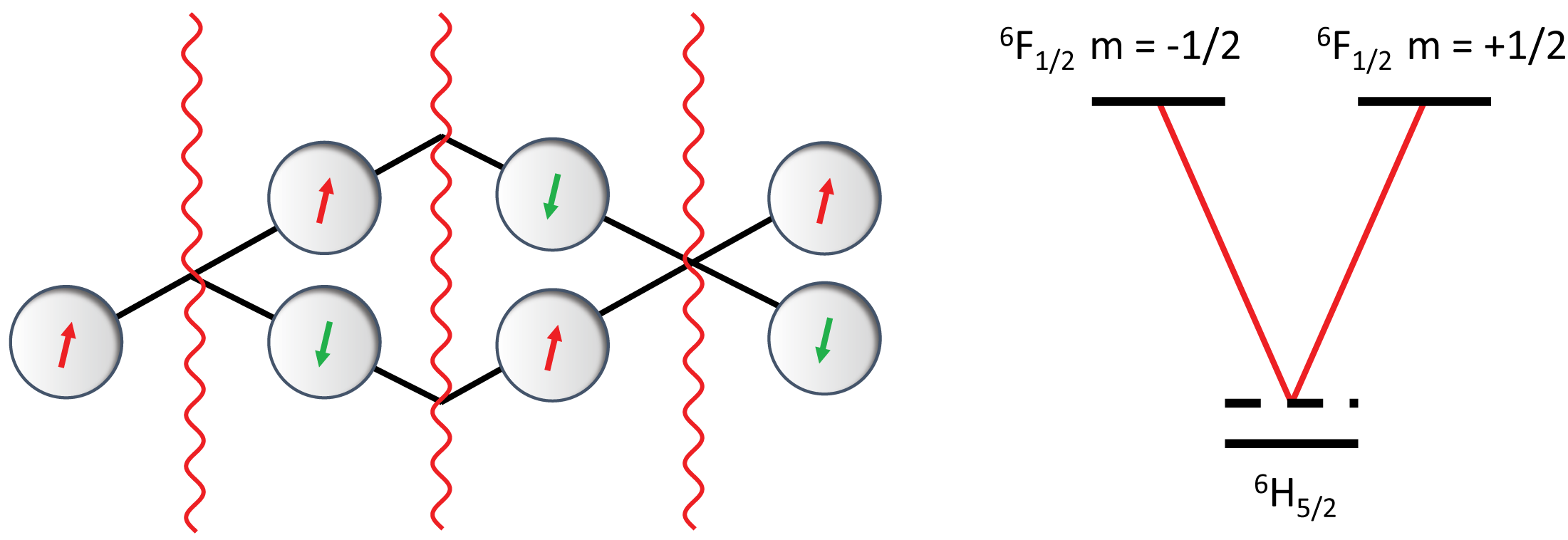}
    \caption{{Experimental concept.  Left:  Mach-Zehnder interferometry sequence of a  microparticle with embedded rare-earth ion.  The rare-earth ion interacts with three optical pulses.  
    The interferometer phase, encoded in the state of the rare-earth ion after the interferometer, is sensitive to the acceleration of the microparticle.    
    Right:  diagram of Raman transitions used to implement the interferometer beam splitter and mirror pulses.  
    The rare-earth ion remains in a state with total angular momentum $J=1/2$ throughout the interferometer, which decouples the state of the rare-earth ion from the microparticle's orientation.
    }} 
    \label{fig:1}
\end{figure}

In principle, this system fulfills all necessary criteria to observe matter-wave interference with a large particle. 

1.  \textit{Beam splitter coherence.}  Matter-wave beam splitters based on optical transitions are widely used in atom interferometry \cite{Kasevich1991} and have demonstrated fringe visibility of $99\%$ with two-photon beam splitters \cite{zhu2026high} and $18\%$ with 600-photon beam splitters \cite{rodzinka2024optimal}.  The use of near-resonant optical transitions to implement the beam splitter pulses allows the momentum transfer to be independent of the presence of defects in the microparticle (e.g., unpaired spins or color centers), as long as these defects do not have resonances at the same frequencies as the optical fields.  Crucially, this relaxes requirements on the surface quality and defect concentration of the microparticle compared to other approaches, e.g. magnetic field gradients \cite{bose2025}.  

2.  \textit{Insensitivity to microparticle state.} The use of a Mach-Zehnder interferometer geometry eliminates the dependence of the interferometer phase on the initial center-of-mass position and momentum of the microparticle, so quantum control of these degrees of freedom is not necessary to observe interference.  In particular, the microparticle need not be cooled to the quantum ground state \cite{delic2020cooling}.  In addition, the use of a quantum state with $J = 1/2$ to implement the beam splitter and the state readout decouples these processes from the orientation of the microparticle.  Although the rare-earth ion is subjected to the internal electric field of the microparticle, which rotates along with the microparticle, the Wigner-Eckart theorem forbids coupling of states within a $J = 1/2$ manifold by any electric field multipole moment.  Thus, the rotational motion of the particle does not decohere the interferometer, and the orientation of the microparticle need not be controlled to observe interference \cite{footnote2}.  Finally, the energy gap between $\pm m$ states is decoupled from electric fields by Kramers' theorem, which prevents elastic phonon scattering from decohering the subspace of these states.  Decoherence induced by magnetic fields can be suppressed by using a material that lacks unpaired electronic and nuclear spins, and the influence of magnetic field noise and magnetic defects can be reduced by using standard spin-echo techniques \cite{souza2012effects}.

3.  \textit{Avoidance of external decoherence.} The microparticle mass is large enough that $\Delta x$ is small compared to the reduced de Broglie wavelength $\lambda$ of any particle that could be scattered by the microparticle, even at scientifically relevant acceleration sensitivities.  In this regime, all decoherence rates are reduced by a factor of $(\Delta x / \lambda)^2$ \cite{decoherence}, and decoherence rates typically decrease (or in the worst case, remain constant) with increasing particle mass.  As we will show in Section~\hyperref[sec:Decoherence]{IV}, for large enough microparticle mass, external decoherence mechanisms are not expected to limit the coherence time even if the system is at moderate vacuum levels.  

\section{III.  Implementation}
\label{sec:Implementation}

\subsection{Rare-earth ion and science states}

No rare-earth ion in the $3+$ charge state has a ground state with total angular momentum $1/2$.  Therefore, the choice of rare-earth ion is determined by the availability of $J = 1/2$ science states with sufficiently long lifetime.  In this work, we consider Sm$^{3+}$, which has a $^{6}F_{1/2}$ excited level and several isotopes with zero nuclear spin.  The interferometer beam splitters can be implemented with two-photon Raman transitions between the $^{6}F_{1/2}$ $m~=~+1/2$ and $m~=~-1/2$ states, encoding the phase of the interferometer in the populations of these states after the final beam splitter.  In solid-state systems, the $^{6}F_{1/2}$ level has a radiative lifetime of about $1.5$~ms and a transition frequency of about $1586$~nm to the electronic ground state \cite{carnall1968electronic}, which can be used as an intermediate state for the Raman transitions.  Using Judd-Ofelt theory \cite{judd1962optical,ofelt1962intensities}, we estimate that the two-photon transition can be driven with $12$~MHz Rabi frequency (see Appendix A.I).

An alternative candidate for the rare-earth ion is Nd$^{3+}$, which has a $^2P_{1/2}$ excited level with a transition frequency of about $431$~nm to the ground state \cite{carnall1968electronic}.  

\subsection{Microparticle material}

The choice of microparticle material is determined by three considerations.  First, the maximum phonon frequency must be low enough to avoid non-radiative decay of the rare-earth ion science states through multi-phonon emission \cite{schuurmans1984radiative} (see Appendix A.I).  In silica, for example, non-radiative decay likely limits the lifetime of the Sm$^{3+}$ $^{6}F_{1/2}$ level to the microsecond time scale.  To suppress non-radiative decay sufficiently to allow a millisecond excited state lifetime, the maximum phonon frequency in the material should be $\lesssim 7$~THz.  This consideration favors the use of materials with heavy atomic constituents.  Second, the band gap of the material must be large enough to accommodate the electronic transitions used for interferometry and imaging ($\sim 3$~eV, for the imaging scheme discussed below).  Third, the density of electronic and nuclear spins in the material must be low enough to permit a millisecond electron spin coherence time.  

Several chalcogenides meet these criteria, including SrSe and BaSe \cite{rakesh2022anomalous}.  The calculations in this work assume that SrSe is used.  Luminescence from Sm$^{3+}$ in related materials such as SrS has been demonstrated previously \cite{yamashita1987photoluminescence}.  Low concentrations of rare-earth ions can be doped into such materials during crystal growth or implanted via ion implantation \cite{rashid2026nitrogen}.  The material can be co-doped with a similar concentration of an alkali atom such as Na or K to break the inversion symmetry of the Sm$^{3+}$ site and to provide charge compensation.

\subsection{Imaging scheme}

The state of the rare-earth ion can be imaged by selectively exciting the $^{6}F_{1/2}$ $m = 1/2$ or $m = -1/2$ state to a higher electronic state that decays radiatively.  One possible approach for samarium, illustrated in Fig. 2, utilizes the $^{6}P_{5/2}$ level.  This level has a $66\%$ radiative decay probability in SrSe and a non-radiative decay pathway to the metastable $^{4}G_{5/2}$ level.  Repumping the $^{4}G_{5/2}$ level and the dominant radiative decay channels of the $^{6}P_{5/2}$ level allows a single ion to scatter $\sim 30$ photons in $1.5$~ms, which is sufficient to distinguish the two magnetic states of the $^{6}F_{1/2}$ level assuming a photon collection efficiency of a few percent.  Table I lists the main decay pathways of the $^{6}P_{5/2}$ level, along with the corresponding rates.  

\begin{figure}[h!]
    \centering
    \includegraphics[width=0.475\textwidth]{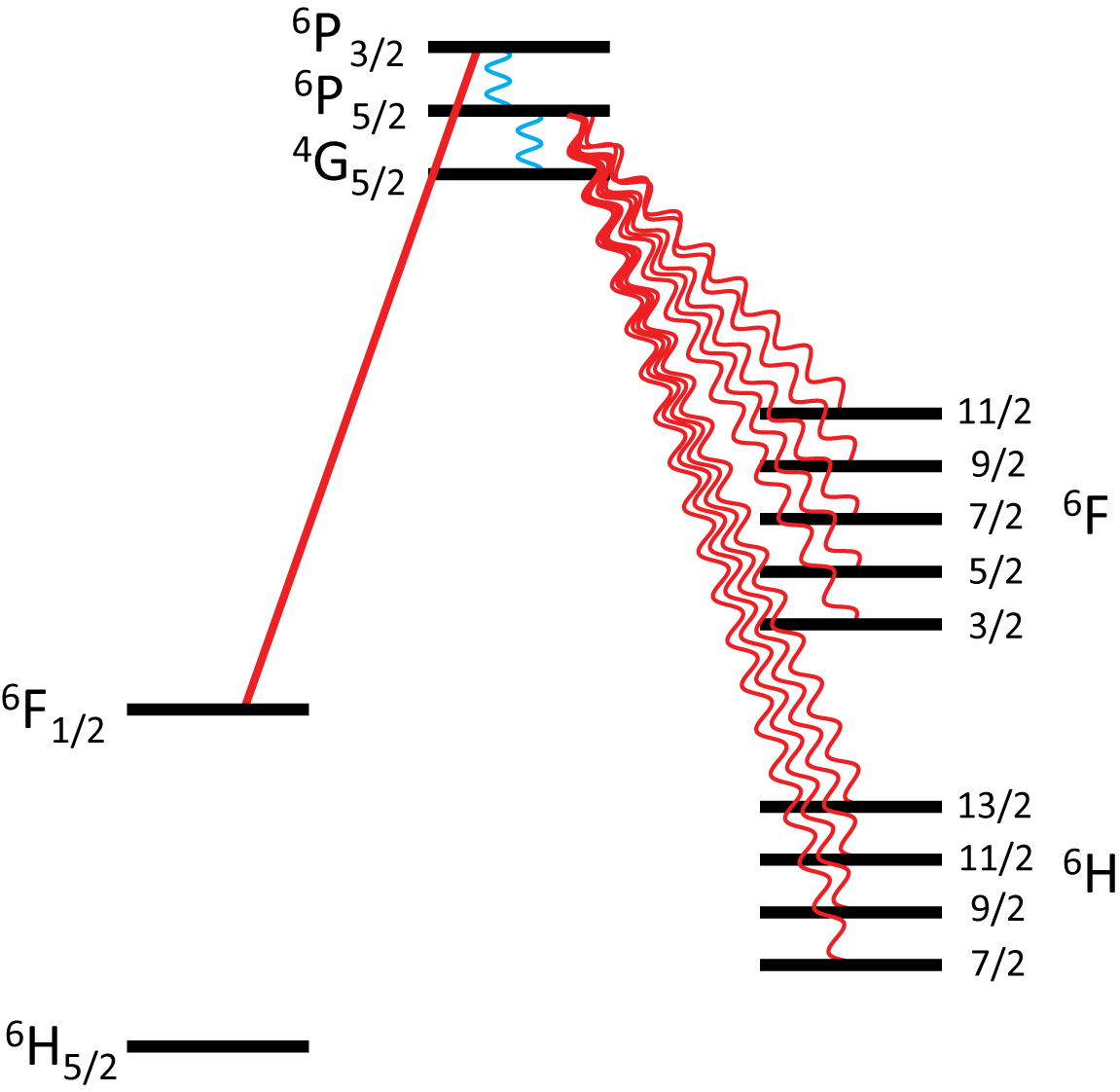}
    \caption{Imaging scheme.  The samarium ion is initially pumped to the $^{6}P_{3/2}$ level, which decays non-radiatively to the $^{6}P_{5/2}$ level.  This level decays non-radiatively to the $^{4}G_{5/2}$ level and radiatively to levels in the $^{6}F$ and $^{6}H$ manifolds.  These levels can be repumped via $^{6}P_{3/2}$ to enable optical cycling.  Figure is not to scale.    
    } 
    \label{fig:2}
\end{figure}

\begingroup
\setlength{\tabcolsep}{8pt} 
\begin{table}[!htbp]
\small
\begin{center}
\begin{tabular}{c|c|c|c}
	\hline  \textbf{Final state} &  \textbf{Type} & \textbf{Rate (s$^{-1}$)} &  \textbf{Ratio} \\

	\hline 
      $^{4}G_{5/2}$ & NR & $1.6 \times 10^4$ & 34\% \\
      $^{6}F_{7/2}$ & ED & $5.6 \times 10^3$ & 12\% \\
      $^{6}F_{9/2}$ & ED & $5.0 \times 10^3$ & 11\% \\
      $^{6}F_{5/2}$ & ED & $4.0 \times 10^3$ & 8\% \\
      $^{6}H_{11/2}$ & ED & $3.7 \times 10^3$ & 8\% \\
      $^{6}H_{9/2}$ & ED & $3.4 \times 10^3$ & 7\% \\
      $^{6}H_{13/2}$ & ED & $2.9 \times 10^3$ & 6\% \\
      $^{6}H_{7/2}$ & ED & $2.1 \times 10^3$ & 4\% \\
      $^{6}F_{3/2}$ & ED & $1.6 \times 10^3$ & 3\% \\
      $^{6}F_{11/2}$ & ED & $9.2 \times 10^2$ & 2\% \\

	\hline 
\end{tabular}
\caption{Calculated decay pathways of $^{6}P_{5/2}$ level of Sm$^{3+}$ in SrSe.  NR:  non-radiative (multi-phonon) transitions.  ED:  electric dipole transition.  About ten repump frequencies are required to optically cycle 30 photons using the $^{6}P_{5/2}$ level.}
\end{center}
\end{table}
\endgroup 

\subsection{Microparticle trap}

To utilize the matter-wave interferometer as an accelerometer while minimizing decoherence rates, the microparticle should be trapped under vacuum, and the trap should be turned off during the interferometer.  
While optical tweezers have been employed to levitate microparticles in vacuum \cite{jin2024towards}, our trap must be nearly dissipationless in order to avoid heating the microparticle.  
Here we propose to trap the microparticle using an AC electric trap \cite{fujio1992electric}.  Such traps, which confine particles at saddle points of a time-dependent electric field through their electric polarizability, have been used to trap neutral atoms \cite{rieger2007trapping}.  Since the polarizability and mass of a particle both scale with its volume, essentially the same trap geometry can be used to levitate a microparticle.  A trap drive frequency of $300$~Hz, electrode spacing of $1$~mm, and maximum voltage $7$~kV yields an estimated trap frequency of $50$~Hz, which is sufficient to support the particle against gravity.

Although the rare-earth ion states used for interferometry are decoupled from the microparticle orientation, the intermediate state used for the Raman transitions, which must have an electric dipole transition to the $J = 1/2$ level to allow sufficiently high Rabi frequency, is necessarily a state with $J > 1/2$.  To ensure that the Rabi frequency of the Raman transitions remains constant from shot to shot, the microparticle can be rotated rapidly enough to fix its axis of rotation in the lab frame, and the rotation axis can be aligned with the direction of the interferometry beams and with the magnetic field.  This configuration also reduces systematic effects associated with the particle's rotation.  Microparticle rotation at a controlled frequency can be induced by applying rotating electric fields that couple to the particle's electric dipole and quadrupole moments \cite{blakemore2022librational,perdriat2024rotational}.  
In addition, since the optical transition frequency of a rare-earth ion likely depends on its depth in the microparticle, the interferometry lasers can be tuned to address a rare-earth ion that is near the center of the microparticle.

To suppress decoherence of the optical transition due to interactions with phonons, the microparticle will be cryogenically cooled by helium buffer gas.  As we show in Section~\hyperref[sec:Decoherence]{IV} and Appendix A.II, the buffer gas pressure can be chosen to maintain the microparticle at cryogenic temperatures without decohering the interferometer.

\subsection{Comparison to atom interferometry}

In addition to increasing the interfering particle mass by many orders of magnitude, matter-wave interferometry with microparticles offers several technical advantages compared to atom interferometry.  The much smaller wave packet separation of the microparticle reduces the length scale over which optical and magnetic fields must be controlled, potentially simplifying precision measurements and enabling more compact sensors.  In addition, the small transverse size of the microparticle allows the interferometry laser to be focused to a much smaller size than is typical in atom interferometry experiments, enabling high Rabi frequency with modest laser powers.  Finally, the use of microparticles rather than atoms eliminates the complexity of preparing an ultracold atom cloud for interferometry, which typically involves magneto-optical trapping, evaporative cooling, and magnetic or optical lensing \cite{Kovachy2015}.  

\section{IV.  Decoherence rates}
\label{sec:Decoherence}

As the radius and geometric cross section of a particle increase, the rate of any scattering process involving the particle generically increases as well.  This might make it seem that scattering-induced decoherence imposes a size limit on the class of particles for which matter-wave interference can be observed.   
However, if the mass of the interfering particle in a Mach-Zehnder interferometer is increased to the point that the wave packet separation $\Delta x$ becomes smaller than the reduced de Broglie wavelength $\lambda$ corresponding to the momentum exchanged with the scattered particle, then the decoherence rate is reduced by a factor of $(\Delta x / \lambda)^2$ (see Appendix A.III).  Physically, when $\Delta x$ becomes smaller than $\lambda$, the scattered particle does not have enough resolution to localize the interfering particle onto one interferometer arm rather than the other, and a single scattering event no longer fully decoheres the interferometer.  

In this ``long-wavelength'' regime where $\Delta x < \lambda$, the decoherence rate of a Mach-Zehnder interferometer due to a scattering process that scales with the geometric cross section of the particle actually decreases with increasing particle mass.  At constant density, the geometric cross section of the particle is proportional to $m^{2/3}$, but $\Delta x \propto m^{-1}$, so the decoherence rate is proportional to $m^{-4/3}$.  Analogously, a scattering process with a rate proportional to the square of the particle volume (e.g., blackbody radiation scattering) leads to a decoherence rate that is constant as a function of particle mass.

In the remainder of this section, we calculate decoherence rates of a Mach-Zehnder microparticle interferometer due to background gas scattering, blackbody radiation, and optical photon scattering. For sufficiently large particles, these  decoherence mechanisms are not expected to prevent the observation of matter-wave interference, even at modest vacuum levels.  Table II lists the properties of the microparticle and of the apparatus that are used for these calculations. 

\begingroup
\setlength{\tabcolsep}{8pt} 
\begin{table}[!htbp]
\small
\begin{center}
\begin{tabular}{c|c|c}
	\hline  \textbf{Property} &  \textbf{Symbol} & \textbf{Value} \\

	\hline 
      Microparticle density & $\rho$ & $4540$ kg/m$^3$ \\
      Relative permittivity & $\epsilon$ & $4.8$  \\
      Pressure & $P$ & $3 \times 10^{-6}$ Torr  \\
      Temperature & $T_e$ & $6$ K  \\
      Background gas mass & $m_a$ & $4$~amu  \\
      Background gas velocity & $v_a$ & $180$ m/s  \\
      Laser wavevector & $k$ & $4 \times 10^6$ /m  \\
      Laser waist & $w$ & $5~\mu$m  \\
      Laser photon flux & $n$ & $8 \times 10^{17}$ /s  \\
      Interferometer time & $T$ & $1$~ms  \\

	\hline 
\end{tabular}
\caption{Typical parameters of microparticle and apparatus used to calculate decoherence rates.}
\end{center}
\end{table}
\endgroup

\subsection{Background gas scattering}

The residual gas will consist primarily of helium, which has a reduced de Broglie wavelength $\lambda_\text{He}~=~90$~pm at $6$~K.  The scattering rate of background gas for a microparticle of mass $m$ can be estimated by \cite{kaltenbaek2012macroscopic}
\begin{equation}
    \Gamma_\text{He} = 2\sqrt{6 \pi}\,\frac{ P}{m_a v_a} \left(\frac{3 m}{4 \pi \rho}\right)^{2/3} \label{eq:BG}
\end{equation}
where $P$ is the pressure, $m_a$ and $v_a$ are the mass and typical velocity of the background gas particles, respectively, and $\rho$ is the microparticle density.  

When $\Delta x > \lambda_\text{He}$, a single background gas collision fully decoheres the interferometer, and the decoherence rate due to background gas collisions is equal to $\Gamma_\text{He}$.  When $\Delta x < \lambda_\text{He}$, the decoherence rate is suppressed by a factor of $(\Delta x / \lambda_\text{He})^2$, and the decoherence rate decreases as a function of microparticle mass. 

\begin{figure}[h!]
    \centering
    \includegraphics[width=0.475\textwidth]{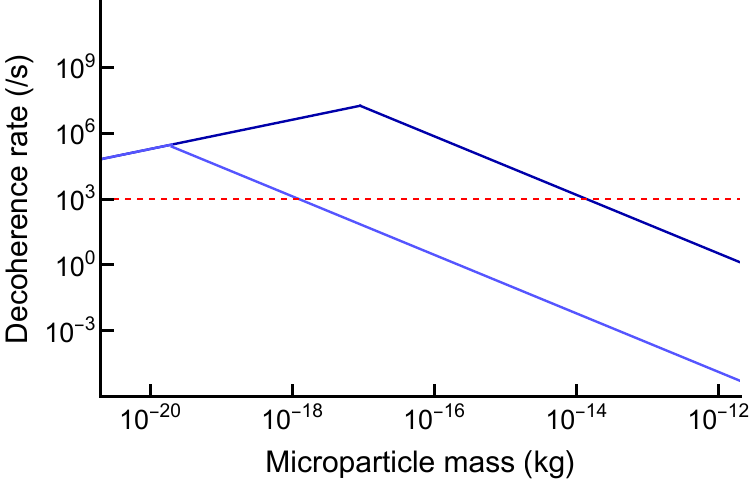}
    \caption{{Estimated decoherence rate of $2\hbar k$ interferometer (light blue) and $1000\hbar k$ interferometer (dark blue) due to background gas scattering as a function of microparticle mass, assuming $3 \times10^{-6}$~Torr pressure at $6$~K.  Dashed red line:  decoherence rate of $10^3$~s$^{-1}$, the maximum rate at which coherence can be observed after 1~ms.       
    }} 
    \label{fig:3}
\end{figure}

Fig. 3 shows the estimated decoherence rate due to background gas scattering in an apparatus with buffer gas pressure $3 \times 10^{-6}$ Torr as a function of particle mass for $2\hbar k$ and $1000\hbar k$ beam splitter momentum.  The interferometer time is chosen to be $T = 1$~ms.  The decoherence rate of each interferometer has a local maximum where $\Delta x \approx \lambda_\text{He}$.  For sufficiently massive particles ($m > 2 \times10^{-14}$~kg, corresponding to a microparticle of diameter $2~\mu$m), this decoherence rate is slow enough to permit the observation of interference with $1000\hbar k$ beam splitters.

\subsection{Blackbody radiation}

At finite temperature, the microparticle interferometer is exposed to blackbody radiation.  The decoherence rate of the interferometer due to blackbody radiation scattering can be estimated by \cite{kaltenbaek2012macroscopic}
\begin{equation}
    F_\text{BBR} = \Lambda_\text{BBR}\, (\Delta x)^2 \label{eq:BBR}
\end{equation}
where
\begin{equation}
    \Lambda_\text{BBR} = \frac{8\cdot 8! \cdot\zeta(9)}{9 \pi}\,c\, \left(\frac{3 m}{4 \pi \rho}\right)^{2} \left(\frac{k_B T_\text{BBR}}{\hbar c} \right)^9 \text{Re} \left(\frac{\epsilon - 1}{\epsilon+2} \right)^2.
\end{equation}
Here $\zeta$ is the Riemann zeta function, $m$ and $\rho$ are the microparticle mass and density, respectively, $T_\text{BBR}$ is the blackbody radiation temperature, and $\epsilon$ is the relative permittivity of the microparticle.  We note that this decoherence rate is constant as a function of mass because the induced dipole of the microparticle scales with its volume.  A conservative estimate using $T_\text{BBR} = 300$~K yields a decoherence rate of $10^{-5}$ s$^{-1}$ for a $1000\hbar k$ interferometer with $1$~ms coherence time.  Thus, blackbody radiation scattering induces negligible decoherence.

We also consider the absorption and emission of blackbody radiation photons by the microparticle.  Decoherence rates due to these processes can be estimated by replacing $\Lambda_\text{BBR}$ in Eq.~\ref{eq:BBR} with the scattering parameter \cite{kaltenbaek2012macroscopic}
\begin{equation}
    \Lambda_\text{BBR}' = \frac{16 \pi^5}{189} c \left(\frac{3 m}{4 \pi \rho}\right)\left(\frac{k_B T_*}{\hbar c} \right)^6 \text{Im} \left(\frac{\epsilon - 1}{\epsilon+2} \right)^2
\end{equation}
where $T_* = T_\text{BBR}$ for absorption or $T_* = T_e$ for emission.   
These decoherence rates, which decrease with increasing particle mass, are many orders of magnitude smaller than that of blackbody radiation scattering and are thus negligible as well.

\subsection{Optical photon scattering}

Optical photons scattered by the microparticle during the beam splitter and mirror pulses can decohere the interferometer.  This effect is most significant in the mirror pulses because the wave packet separation is largest there.  The decoherence rate due to optical photon scattering is approximately
\begin{equation}
    F_\text{laser} = \Lambda_\text{laser}\, (\Delta x)^2
\end{equation}
where
\begin{equation}
    \Lambda_\text{laser} = n\, f\, \left(\frac{3 m}{4 \pi \rho}\right)^{2/3} w^{-2}\, k^2.
\end{equation}
Here $n$ is the incident number of photons per second, $m$ and $\rho$ are the microparticle mass and density, respectively, $w$ is the laser waist, $k$ is the laser wavevector, and $f$ is a dimensionless number that characterizes the scattering efficiency.   

\begin{figure}[h!]
    \centering
    \includegraphics[width=0.475\textwidth]{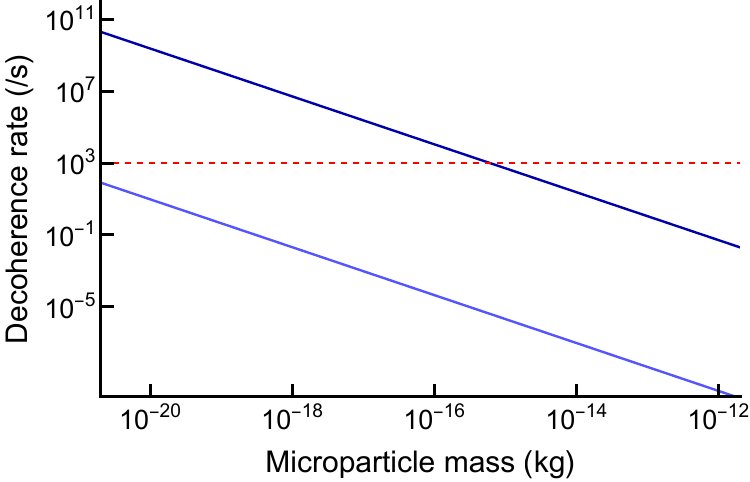}
    \caption{{Estimated decoherence rate of $2\hbar k$ interferometer (light blue) and $1000\hbar k$ interferometer (dark blue) due to optical photon scattering as a function of microparticle mass.  Dashed red line:  decoherence rate of $10^3$~s$^{-1}$, the maximum rate at which coherence can be observed after 1~ms.       
    }} 
    \label{fig:4}
\end{figure}

Fig. 4 shows the decoherence rate due to optical photon scattering for a $T = 1$~ms interferometer as a function of particle mass, using the parameters listed in Table II.  In general, $f$ depends on the particle radius and thus on the particle mass.  Numerical simulations indicate that $f$ varies from about $0.7$ to $1.3$ for particles in the Mie regime.  For simplicity, Fig. 4 is plotted with $f = 1$.  The decoherence rate is time averaged and takes into account the total duration of the mirror pulses for each interferometer, assuming that the two-photon Rabi frequency is $12$~MHz (see Appendix I).  For $m > 10^{-15}$~kg, the decoherence rate is low enough to permit observation of interference in a $1000\hbar k$ interferometer after $1$~ms.  

If the rate of momentum transfer is limited by the Rabi frequency of the Raman transitions, this decoherence rate is proportional to the cube of the applied optical power.  
Optical photon scattering is the dominant external decoherence mechanism for sufficiently high optical power.

\section{V.  Discussion}
\label{sec:Discussion}

Matter-wave interferometry with microparticles will be immediately impactful in tests of minimal modifications of quantum mechanics \cite{nimmrichter2013macroscopicity}.  Such modifications, which include the continuous spontaneous localization model \cite{ghirardi1990markov}, can be constrained by observing coherence in spatial superposition states of massive particles.  The constraints from a given experiment can be quantified by a macroscopicity parameter \cite{nimmrichter2013macroscopicity}.  A first-generation microparticle interferometry experiment could achieve macroscopicity $\mu \approx 18.5$ at a length scale of about $20$~fm, which would improve by three orders of magnitude over the current state of the art \cite{pedalino2026probing}.

With $10$~ms cycle time and a single microparticle per shot, a first-generation microparticle interferometer could reach an acceleration sensitivity of $10^{-5}\;g/\sqrt{\text{Hz}}$ and could find applications as a compact sensor for accelerometry \cite{Peters2001}, rotation sensing \cite{Gustavson1997}, and gravity gradiometry \cite{McGuirk2002}.  This system could also be used for short-range tests of gravity \cite{murata2026}.  

To reach state-of-the-art acceleration sensitivity for tests of the equivalence principle or searches for dark matter, the coherence time would need to be extended.  One possibility is to dope the SrSe microparticle with $^{125}$Te, which substitutes for Se$^{2-}$ and has $I = 1/2$.  After the initial beam splitter sequence, the quantum state of the rare-earth ion could be transferred to the nuclear spin of a nearby $^{125}$Te atom.  This process would be time-reversed to allow subsequent beam splitter pulses to be implemented.  Quantum state transfer from electrons to nucleons has previously been demonstrated \cite{aiello2026}, and nuclear spins in cryogenic solid-state systems have demonstrated coherence times of up to six hours \cite{zhong2015optically}.  Sensitivity could also be increased by loading many microparticles into the trap and performing interferometry in parallel.

Although the use of an optical beam splitter is favorable for initial observations of coherence and for accelerometry, decoherence from optical photon scattering will prevent this technique from being extended to wave packet separations that would allow tests of gravitationally mediated entanglement generation.  Nevertheless, the approach outlined in this work will pave the way for such experiments by allowing the decoherence mechanisms of microparticle interferometers to be studied and minimized.  In place of an optical beam splitter, one could imagine using a beam splitter based on magnetic field gradients \cite{bose2025} or a hybrid beam splitter that avoids applying optical fields when the wave packet separation is large.

\section{Acknowledgments}

We thank David DeMille, Alexander O. Sushkov, Marcus Aspelmeyer, Michael Hesford, David E. Kaplan, Tyrel M. McQueen, Surjeet Rajendran, and Mohit Verma for useful discussions.  

\section{Appendix}
\subsection{A.I.  Judd-Ofelt theory of Sm$^{3+}$ in SrSe}

In a free rare-earth ion, electric dipole transitions between $4f$ states are forbidden by the parity selection rule.  When the ion is embedded in a solid-state material, however, the electric field of the material weakly mixes the $4f$ states with $5d$ states, allowing electric dipole transitions to occur.  The matrix elements of such transitions can be calculated with Judd-Ofelt theory \cite{judd1962optical,ofelt1962intensities}, which relates the magnitude of a given matrix element to three material-dependent parameters $\Omega_2$, $\Omega_4$, and $\Omega_6$.  In principle, these parameters can be calculated from electric field multipole moments at the location of the rare-earth ion; in practice, Judd-Ofelt parameters are determined empirically by fitting to the observed line strengths of a set of rare-earth ion optical transitions. 

The radiative decay rate of the $^{6}F_{1/2}$ level can be calculated by summing over its radiative decay rates to each lower energy level.  The oscillator strength of the manifold-to-manifold transition from level $\ket{\phi_i}$ to level $\ket{\phi_f}$ is given by \cite{walsh2006judd}
\begin{equation} \label{eq:Osc}
    f = \frac{8\pi^2 mc}{3h\lambda(2J+1)} n \left(\frac{n^2+2}{3n}\right)^2 \sum_{i=2,4,6} \Omega_i |\bra{\phi_f}|\textbf{U}^{(i)}| \ket{\phi_i}|^2.
\end{equation}

Here $m$ is the electron mass, $\lambda$ is the transition wavelength, $J$ is the total angular momentum of the initial level, $n$ is the index of refraction, the $\textbf{U}^{(i)}$ are irreducible tensor operators of rank $i$ that are constructed by combining odd-order multipole moments of the crystal electric field with the electric dipole operator, and the $\bra{\phi_f}|\textbf{U}^{(i)}| \ket{\phi_i}$ are doubly reduced matrix elements.

The values of reduced matrix elements were obtained from the AMELI repository \cite{caspary2026ameli} using the YALIP application software.  To our knowledge, the Judd-Ofelt parameters of Sm$^{3+}$ in SrSe have not previously been measured.  From Judd-Ofelt analyses of Sm$^{3+}$ in selenide-chalcogenide glasses \cite{starecki20187,nvemec2006optical}, we estimate the parameters $\{\Omega_2, \Omega_4, \Omega_6\}$ = $\{3, 2,1.5\} \times 10^{-20}$ cm$^2$ for Sm$^{3+}$ in SrSe.  The estimated radiative decay rate of the $^{6}F_{1/2}$ level is calculated to be $7.5 \times 10^2$~s$^{-1}$, which implies a lifetime of $1.3$~ms if the decay is radiatively dominated.  

Non-radiative decay of excited states occurs via multi-phonon emission at the rate 
\begin{equation}
    W \approx B \exp (-\beta\, \Delta E)
\end{equation}
where $\Delta E$ is the energy gap to the next lower energy level.  We estimate the parameters $B$ and $\beta$ from measurements of non-radiative decay rates in KPb$_2$Cl$_5$ \cite{nostrand2001optical}, which has a similar maximum phonon frequency.  Using $B = 3.72 \times 10^9$ s$^{-1}$ and $\beta = 1.16 \times 10^{-2}$~cm, we obtain non-radiative decay rates of $1.2 \times 10^3$~s$^{-1}$ for the $^{6}F_{1/2}$ level at room temperature and $4\times10^1$~s$^{-1}$ at cryogenic temperatures.  The estimated non-radiative decay rate at cryogenic temperatures is significantly smaller than the estimated radiative decay rate.   

The Rabi frequency of a given transition can be calculated from the oscillator strength (Eq.~\ref{eq:Osc}).  For the $^{6}H_{5/2}$~-~$^{6}F_{1/2}$ transition to be used for interferometry, we obtain single-photon Rabi frequency $\Omega = 1.5 \times 10^6\, \sqrt{I/( \text{W/cm}^2)}$ rad/s for laser intensity $I$.  With $100$~mW focused to a waist of $5~\mu$m and a single-photon detuning of $650$~MHz, the two-photon Rabi frequency is $12$~MHz.  These parameters allow sufficient Rabi frequency to implement a $1000\hbar k$ interferometer with $T = 1$~ms while keeping the transient population of the intermediate state below $1\%$ to reduce systematic effects.  With excellent control of pulse fidelity errors, interferometry with up to $10^4$ photons per beam splitter may be possible in this system.

\subsection{A.II.  Buffer gas cooling of a microparticle}

Buffer gas cooling of levitated microparticles has been demonstrated previously \cite{jin2024towards}.  To estimate the cooling rate at a given buffer gas temperature and pressure, we assume that each scattering event removes $k_B~(T_\text{mp}-T_\text{gas})$ of energy from the microparticle, where $T_\text{mp}$ is the microparticle temperature and $T_\text{gas}$ is the background gas temperature.  The cooling rate can then be calculated from the background gas scattering rate (Eq.~\ref{eq:BG}) and the specific heat of the microparticle [$280$~J/(kg K) at room temperature for SrSe, decreasing to about $0.4$ J/(kg K) at $10$~K].

A helium buffer gas pressure of $10^{-4}$ Torr at $6$~K is sufficient to cool a microparticle of $10~\mu$m diameter from room temperature to $<10$~K in a few seconds.  Once the microparticle is loaded into the AC electric trap and cooled, the background gas pressure can be reduced to $3 \times 10^{-6}$ Torr, which is sufficient to maintain the microparticle temperature at $\sim15$~K with $10$~ms cycle time.  We note that these buffer gas pressures are low enough to be compatible with the large electric fields created by the AC electric trap \cite{irmisch1993breakdown}.  

The dominant sources of heating during the measurement sequence are anticipated to be absorption of light from the interferometry lasers and during imaging, which can heat a microparticle of $10~\mu$m diameter by $\sim10^{-13}$~J per shot.  Dissipation induced by the AC electric trap is expected to be negligible.

\subsection{A.III.  Model of decoherence}

Following Ref. \cite{decoherence}, we model a scattering process between the microparticle system $S$ and a single particle in the environment $E$ by assuming that the system and environment are initially uncorrelated.  The initial density matrix can be written as 
\begin{equation}
    \rho(0) = \rho_S(0) \otimes \rho_E (0)
\end{equation}
where $\rho_S$ and $\rho_E$ are the reduced density matrices of the system and environment, respectively.  In the position basis, the density matrix can be expressed as 
\begin{equation}
    \rho(0) = \int dx \int dx'\, \rho_S(x, x', 0) \ket{x}\bra{x'}\otimes\ket{\chi_i}\bra{\chi_i} 
\end{equation}
where $\ket{x}$ is a center-of-mass position eigenstate of the microparticle and $\ket{\chi_i}$ is the initial state of the environmental particle.  Assuming that the scattering interaction is invariant under a joint translation of $S$ and $E$, and neglecting the recoil of the microparticle due to the scattering event, the density matrix after the scattering event is given by 
\begin{equation}
    \rho = \int dx \int dx'\, \rho_S(x, x', 0) \ket{x}\bra{x'}\otimes\ket{\chi(x')}\bra{\chi(x)} 
\end{equation}
where $\ket{\chi(x)}$ is the final state of the environmental particle scattered at position $x$.  The reduced density matrix of $S$ is then given by 
\begin{align}
    \rho_S &= \text{Tr}_E\, \rho \nonumber \\ &= \int dx \int dx'\, \rho_S(x, x', 0) \ket{x}\bra{x'} \braket{\chi(x')| \chi(x)}. 
\end{align}
Therefore, the effect of the scattering event is to transform $\rho_S(x,x',0) \rightarrow \rho_S(x,x',0) \braket{\chi(x')| \chi(x)}$, and the loss of spatial coherence of $S$ after the scattering event is determined by the degree of overlap of the final states of the environmental particle.  

As in Ref. \cite{decoherence}, if we now consider a sequence of many scattering events, the off-diagonal elements of the reduced density matrix decay in time according to   
\begin{equation}
    \frac{\partial \rho_S (x,x',t)}{\partial t} = -F(x - x') \rho_S(x,x',t).
\end{equation}
If we assume that the environmental particles are isotropically distributed, the decoherence rate $F$ is given by
\begin{align}
    &F(x - x') = \nonumber \\ &\int dq\, p(q)\, \nu(q) \int \frac{d\hat{n}\, d\hat{n}'}{4 \pi} (1 - e^{iq (\hat{n} - \hat{n}')\cdot(x - x')/\hbar}) |f(q\hat{n}, q\hat{n}')|^2. \label{eq:F}
\end{align}
Here $p(q)$ is the momentum distribution of environmental particles, normalized so that $\int dq\, p(q) = N/V$ for total particle number $N$ and volume $V$; $\nu(q)$ is the speed of an environmental particle with momentum $q$; $\hat{n}$ and $\hat{n}'$ are unit vectors in momentum space; $d\hat{n}$ and $d\hat{n}'$ are solid-angle differentials in momentum space; and $|f(q_f, q_i)|^2$ is the differential cross section for scattering a particle from initial momentum $q_i$ to final momentum $q_f$. 

We consider a system in an initial spatial superposition state between two positions separated by a distance $\Delta x = |x - x'|$.  The qualitative behavior of $F(x - x')$ is determined by whether the typical reduced de Broglie wavelength of the momentum exchange $\lambda = \hbar / q$ is small or large compared to $\Delta x$.  When $\lambda \ll \Delta x$, the exponential term in the integrand of Eq.~\ref{eq:F} oscillates rapidly and does not contribute significantly to the integral.  In this ``short wavelength'' limit, the decoherence rate simplifies to \cite{decoherence}
\begin{equation}
    F(x - x') \rightarrow \int dq\, p(q) \nu(q) \sigma(q) = \Gamma
\end{equation}
where $\sigma(q)$ is the total cross section and $\Gamma$ is the total scattering rate.  In this regime, a single scattering event encodes complete ``which-path'' information about the system into the environment, so a single scattering event produces complete spatial decoherence.  

On the other hand, in the ``long-wavelength'' limit where $\lambda \gg \Delta x$, the exponential term in the integrand of Eq.~\ref{eq:F} is close to unity.  In this limit, the decoherence rate simplifies to \cite{decoherence}
\begin{equation}
    F(x - x')\rightarrow \Lambda\, (x - x')^2
\end{equation}
where
\begin{equation}
    \Lambda = \int dq\, p(q)\, \nu(q)\, \frac{q^2}{\hbar^2}\, \sigma_\text{eff}(q)
\end{equation}
and $\sigma_\text{eff}(q)$ is an effective scattering cross section, which can be approximated by $\sigma(q)$.  In this regime, each scattering event encodes partial ``which-path'' information about the system into the environment, and many scattering events are required to fully spatially decohere the system.  Compared to the short-wavelength limit, the decoherence rate in the long-wavelength limit is suppressed by a factor of $(\Delta x/\lambda)^2$. 

To calculate the decoherence rates shown in Figs. 3 and 4, we conservatively treat the interferometer arms as being separated by a constant distance $\Delta x$ throughout the interferometer time.  


\begin{thebibliography}{51}
\bibitem{footnote2}
The crystal field mixes the $J = 1/2$ manifold with states of larger $J$.  The effect of this mixing is to make the g-factor of the $J = 1/2$ states anisotropic.  Rotation of the microparticle in the presence of g-anisotropy introduces a Berry's phase that is echoed out by a Mach-Zehnder interferometry sequence.
\end{thebibliography}

\end{document}